\documentclass[preprint,prd,noshowpacs,nofootinbib]{revtex4}

\usepackage{color} 
\usepackage{amssymb}
\usepackage{amsmath}
\usepackage{graphicx}

\usepackage{pdfpages}
\usepackage{epstopdf}
\usepackage[utf8]{inputenc}

\begin{document}

\title{Modified commutators are not sufficient to determine a quantum gravity minimal length scale}

\author{Michael Bishop}
\email{mibishop@mail.fresnostate.edu}
\affiliation{Mathematics Department, California State University Fresno, Fresno, CA 93740}

\author{Jaeyeong Lee}
\email{yeong0219@mail.fresnostate.edu}
\affiliation{Physics Department, California State University Fresno, Fresno, CA 93740}

\author{Douglas Singleton}
\email{dougs@mail.fresnostate.edu}
\affiliation{Physics Department, California State University Fresno, Fresno, CA 93740}

\date{\today}

\begin{abstract}
    In quantum gravity it is generally thought that a modified commutator of the form $[{\hat x}, {\hat p}] = i \hbar (1 + \beta p^2)$ is sufficient to give rise to a minimum length scale. We test this assumption and find that different pairs of modified operators can lead to the same modified commutator and yet give different or even no minimal length. The conclusion is that the modification of the operators is the main factor in determining whether there is a minimal length. This fact - that it is the specific form of the modified operators which determine the existence or not of a minimal length scale - can be used to keep or reject specific modifications of the position and momentum operators in theory of quantum gravity .        
\end{abstract}

\maketitle

\section{Quantum gravity minimal length scale}

There are generic and broad arguments that any theory of quantum gravity  should have some fixed fundamental minimal length scale \cite{maggiore,amati,amati2,gross,scardigli}. 
As pointed out in \cite{garay}, a fundamental length scale can be obtained generically from many different approaches to quantum gravity ({\it e.g.} string theory, loop quantum gravity, path integral quantum gravity) and would have many important observable consequences for cosmology such as the smoothing of the ultraviolet behavior of particle physics and the small distance structure of the Lorentz transformations. 
At low momentum, the relationship between the uncertainty in position and momentum should take the standard form, namely $\Delta x \Delta p \sim const. \Rightarrow \Delta x \sim \frac{const.}{\Delta p}$. 
At large enough momentum, quantum gravity arguments indicate a linear relationship between $\Delta x$ and $\Delta p$, namely $\Delta x \sim \Delta p$ \cite{maggiore} \cite{scardigli}.
These physically motivated arguments imply the following generalized uncertainty principle (GUP) \cite{adler-1999}
\begin{equation}
\label{GUP}
    \Delta x\Delta p \ge \frac{\hbar}{2}\Bigg\lbrack 1  + \beta {(\Delta p)}^2\Bigg\rbrack ~,
\end{equation}
where $\beta$ is a phenomenological parameter that is assumed to be connected with the scale of quantum gravity. The GUP in \eqref{GUP} leads to the relationship $\Delta x \sim \frac{1}{\Delta p} + \beta \Delta p$ which, as shown in \cite{KMM}, leads to a minimum length of order $\sqrt{\beta}$. Good reviews of the appearance of a minimal length scale in quantum gravity can be found in \cite{piero-2009, hossenfelder}. One important feature of this phenomenological GUP approach to a minimal length scale from quantum gravity is that these models open up the possibility of experimental tests. A recent overview of the status of experimental tests and experiment limits in GUP models can be found in \cite{chang}.

Any uncertainty principle between position and momentum is connected with a commutator between position and momentum operators. 
In reference \cite{KMM}, Kempf, Mangano, and Mann (KMM) connected the GUP in equation \eqref{GUP} with the following modified commutator 
\begin{equation}
    \label{GCOM}
    [{\hat x}', {\hat p}'] = i \hbar (1 + \beta p^2) ~.
\end{equation}
The primes on the position and momentum operators on the left hand side of \eqref{GCOM} indicate that these operators are modified versions of the standard quantum position operator (${\hat x} = i \hbar \partial _p$ in momentum space or ${\hat x} = x$ in coordinate space) and momentum operator (${\hat p} = -i \hbar \partial _x$ in coordinate space or ${\hat p} = p$ in momentum space). 
In this work, primed operators will indicate modified operators and unprimed will be the usual operators. 
On the right hand side of \eqref{GCOM}, the $p^2$ term is the standard quantum momentum and not the modified momentum. In the following section, we show that there are a host of different ways to modify the operators, ${\hat x}'$ and ${\hat p}'$, which yield the exact same modified commutator in equation \eqref{GCOM}. 
However, these different ways to modify the position and momentum operators do not all lead to the GUP in \eqref{GUP} and do not all lead to a minimum length. 
This is the main point of this work -- that it is the modification of the position and momentum operators, more than the modification of the commutator, which determines if there is a minimum length scale.   

\section{Modified Uncertainty Relationship}

For two generic operators $A$ and $B$, the connection between their uncertainties $ \Delta A$ and $\Delta B$ and their commutator $[A, B]$ is given by \cite{weinberg} 
\begin{equation}
\label{UP}
      \Delta A \Delta B \ge \frac{1}{2}\mid \langle \lbrack A,B\rbrack \rangle \mid ~.
\end{equation}
We now apply the relationship in \eqref{UP} to the various operators in the previous section and show that despite all the operators having the same modified commutator, the uncertainty relationship is qualitatively different for the different operators.

We begin by examining the original KMM modified  operators \cite{KMM} where ${\hat x}' = i\hbar (1 + \beta p^2)$ and ${\hat p}' = {\hat p}$,  ({\it i.e.} the momentum operator is not modified). Inserting these into \eqref{UP} yields
\begin{equation}
\label{KMM-UP}
    \Delta  x' \Delta p \ge \frac{\hbar}{2} \left (1 + \beta (\Delta p)^2 + \beta \langle \hat p \rangle ^2 \right) ~,
\end{equation}
where we have used $(\Delta p)^2 = \langle \hat p ^2 \rangle - \langle \hat p \rangle ^2$. 
An important point to note is that we have written $\Delta p$ rather than $\Delta p '$ since the momentum operator just is the standard one. 
This is not the case for all the other modified operators. 
One can write $\Delta x'$ as a function of $\Delta p$ and find that $\Delta x' \sim \frac{1}{\Delta p} + \beta \Delta p$. 
As shown in \cite{KMM}, this leads to a minimal uncertainty in position of $\Delta x _0 = \hbar \sqrt{\beta}$. 

A key point in obtaining this minimal distance, connected with the modified position and momentum operators of \cite{KMM}, is that the momentum operator is simply the regular momentum operator from quantum mechanics. 
Thus, the $\Delta p$ which appears on the left hand side and right hand side of \eqref{KMM-UP} are the same. 
This is what allows one to write a direct lower bound on $\Delta x'$.  
The competition between a decreasing $\frac{1}{\Delta p}$  and an increasing $\Delta p$ term leads a strictly positive lower bound on $\Delta x'$ and a minimum length scale. 
When the modified momentum operator is different from the usual momentum operator then equation \eqref{KMM-UP} is replaced by 
\begin{equation}
\label{KMM-GUP}
    \Delta  x' \Delta p' _p \ge \frac{\hbar}{2} \left (1 + \beta (\Delta p)^2 + \beta \langle \hat p \rangle ^2 \right) ~,
\end{equation}
where $p' _p$ indicates that the $p'$ (the modified momentum  operator) is a function of $p$ (the standard momentum operator). 
This function $\Delta p' _p$ will in general be different for each different modified momentum operator ${\hat p}'$. 
Depending on the specific form of $p' _p$, \eqref{KMM-GUP} may still have a global minimum for $\Delta x '$, or it may have only a local minimum for $\Delta x '$, or may not lead to a minimum for $\Delta x '$ at all. 
For example, if the functional dependence of $\Delta p' _p$ on $\Delta p$ is of the form $\Delta p' _p \propto (\Delta p)^\gamma$ then the relationship in \eqref{KMM-GUP} would yield $\Delta  x' \sim \frac{const.}{(\Delta p)^\gamma} + \frac{\beta}{\Delta p ^{\gamma-2}}$. 
If $0 \le \gamma \le 2$ then there exists a minimum length scale, although for $\gamma =0 $ this minimum scale occurs at $\Delta p=0$, which is unphysical. Thus the existence or not of a minimum length scale depends crucially on the functional form of $\Delta p_p '$ in terms of $\Delta p$, which ultimately depends on the relationship between ${\hat p'}$ and ${\hat p}$. 
In general, \eqref{KMM-GUP} can be used to put a lower bound on $\Delta  x'$ of the form
\begin{equation}
\label{KMM-GUP1}
    \Delta  x'  = \frac{\hbar}{2} \frac{\left (1 + \beta (\Delta p)^2 + \beta \langle \hat p \rangle ^2 \right)}{\Delta p' _p} ~.
\end{equation}

The difficulty in the above approach in finding whether or not there is a minimum length is that it is hard to write down the functional relationship between $\Delta p' _p$ and $\Delta p$, {\it i.e.} to find $\Delta p' _p$ as a function of $\Delta p$. To deal with this issue, we will examine \eqref{KMM-GUP1} for specific states, $| \Psi_{test} \rangle$ in the following section. This will allow the explicit calculations of $\Delta p' _p$ and $\Delta p$. We then investigate whether or not the generalized uncertainty principle (GUP) in equation \eqref{KMM-GUP1} gives a minimum length. 

\section{Determination of minimal lengths for various modified position and momentum operators}

There is a large family of modified operators which all have the commutation relation $[x',p'] = i\hbar(1 + \beta p^2)$.  The modification of the position and momentum operator of reference \cite{KMM} which gave the modified commutator in \eqref{GCOM} was
\begin{equation}
    \label{KMM-xp}
   {\hat x'} = i{\hbar}(1 + \beta p^2) \partial_p  ~~~;~~~
   {\hat p'} = p ~.
\end{equation}
The modified operators in \eqref{KMM-xp} are in momentum space and ${\hat x} = i \hbar \partial _p$ is the usual position operator. 
The momentum operator is not altered from the standard form. 

A simple way to get the position space versions of the momentum space modified operators in \eqref{KMM-xp} is to make the simple change $i \hbar \partial_p \mapsto x$ and $p \mapsto -i \hbar \partial _x$ so that \eqref{KMM-xp} becomes
\begin{equation}
    \label{KMM-xpx}
   {\hat x'} = (1 - \beta \hbar ^2 \partial ^2 _x) x ~~~;~~~
   {\hat p'} = -i \hbar \partial _x ~.
\end{equation}
There is some freedom in \eqref{KMM-xpx} in that one could also write the modified position operator as ${\hat x'} = x (1 - \beta \hbar ^2 \partial ^2 _x)$ {\it i.e.} with $x$ leading the derivative operator. Both \eqref{KMM-xp} and \eqref{KMM-xpx} lead to the same modified commutator as can be seen by explicitly plugging ${\hat x}'$ and ${\hat p}'$, from either \eqref{KMM-xp} or \eqref{KMM-xpx}, into the commutator in \eqref{GCOM}.


Equations \eqref{KMM-xp} \eqref{KMM-xpx} are not the only ways to write down the modified operators. A different set of modified operators which treat the modified position and momentum equally and still lead to the modified commutator in \eqref{GCOM} is given by \cite{BAS}
\begin{equation}
    \label{BAS-xp}
   {\hat x'} = i{\hbar} e^{-\beta p^2/2} \partial_p  ~~~;~~~
   {\hat p'} = e^{\beta p^2/2} p ~.
\end{equation}
One can obtain the position space version of the modified operators in \eqref{BAS-xp} by again making the simple replacement $i \hbar \partial _p \mapsto x$ and $p \mapsto -i \hbar \partial _x$ to give
\begin{equation}
\label{BAS-xp1}
    \hat{x}' = e^{\beta \hbar^2 \partial ^2 _x / 2 }x~, ~~~;~~~ \hat{p}' = -i\hbar~e^{-\beta \hbar^2 \partial ^2 _x / 2 } \partial _x ~.
\end{equation}
Substituting the operators from either \eqref{BAS-xp} or \eqref{BAS-xp1} into $[{\hat x}', {\hat p}']$ leads to the modified commutator in \eqref{GCOM}. There is again an ambiguity in the position operator in \eqref{BAS-xp1} since one could have defined the position operator as $x e^{\beta \hbar^2 \partial ^2 _x / 2 }$ and this would also give the commutator in \eqref{GCOM}. 

As a final example of another variant of modified operators that lead to the same modified commutator in \eqref{GCOM}, we give a variant of the operators from reference \cite{das2} \cite{BDV} which in position space can be written as
\begin{equation}
\label{BDV}
    \hat{x}' = x~, ~~~;~~~ \hat{p}' = -i\hbar \left( 1 - \frac{\beta \hbar ^2}{3} \partial ^2 _x \right) \partial _x ~.
\end{equation}
Again plugging the operators from \eqref{BDV} into $[{\hat x}', {\hat p}']$ leads to \eqref{GCOM}. The modified operators in \eqref{BDV} leave the position operator unchanged, and the momentum operator is the one which is altered. As before the momentum space version of the operators in \eqref{BDV} can be obtained via  $ x \mapsto i \hbar \partial _p$ and $-i \hbar \partial _x \mapsto p$ which gives
\begin{equation}
\label{BDV-1}
    \hat{x}' = i \hbar \partial _p~, ~~~;~~~ \hat{p}' = p + \frac{\beta}{3} p^3  ~.
\end{equation}
Again by plugging the operators from \eqref{BDV-1} into $[{\hat x}', {\hat p}']$ results in the modified commutator in \eqref{GCOM}. Some subtle issues connected with the operators in \eqref{BDV} and \eqref{BDV-1}, with the position operator unmodified, are discussed in \cite{bosso1}. 

In the above equations, we have collected a series of different modified operators (equations \eqref{KMM-xp}, \eqref{BAS-xp}, \eqref{BDV-1} in momentum space, and equations \eqref{KMM-xpx},   \eqref{BAS-xp1}, \eqref{BDV} in position space) which all lead to the same modified commutation relationship \eqref{GCOM}. There are a host of other modified operators that lead to \eqref{GCOM}, but for the point of this work it is enough the consider the modified operators collected above.  

We now look at the modified uncertainty principle obtained for the above modified operators \eqref{BAS-xp}, \eqref{BDV-1}, and \eqref{KMM-xp} in momentum space.  
We will find that even though all the modified operators lead to the same modified commutator in equation \eqref{GCOM}, the uncertainty relationships are all different. 

\subsection{Gaussian modified position and momentum}

Let us begin by examining the ``Gaussian" modified position and momentum operators of \eqref{BAS-xp} of the form ${\hat x}' = i \hbar e^{-\beta p^2 /2} \partial _p$ and ${\hat p}'= e^{\beta p^2 /2} p$. For our test wave function we choose a Gaussian in momentum space
\begin{equation}
    \label{psi-test}
    | \Psi_{test} \rangle = \Psi (p) = C e^{-p^2/2 \sigma ^2} ~.
\end{equation}
To determine the normalization $C$, one normally would calculate $\int \Psi (p) ^* \Psi (p) dp =1$. 
However, as pointed out in reference \cite{KMM}, modifying the position operator to the form ${\hat x}' = i \hbar f(p) \partial _p$ requires modifying the normalization integral as $\int \Psi (p) ^* \Psi (p) \frac{dp}{f(p)} =1$; this change is needed in order to make ${\hat x}'$ symmetric {\it i.e.} to have $(\langle \Psi | {\hat x}' ) | \Phi \rangle  = \langle \Psi | ({\hat x}'  | \Phi \rangle )$. 
Using this modified $p$-space integral related to the modified position operator ${\hat x}' = i \hbar e^{-\beta p^2 /2} \partial _p$ (here $f(p) = e^{-\beta p^2 /2}$), we determine the normalization constant $C$ in \eqref{psi-test} via $\int \Psi (p) ^* \Psi (p) \frac{dp}{f(p)} =1 ~~\Rightarrow~~ |C|^2 \int _{-\infty} ^{\infty} e^{-p^2 /a^2} dp =1 ~~ \Rightarrow ~~ C= \frac{1}{(\sqrt{\pi} a) ^{1/2}}$ which yields 
\begin{equation}
    \label{psi-test1}
    | \Psi_{test} \rangle = \Psi (p) = \frac{1}{(\sqrt{\pi} a) ^{1/2}} e^{-p^2/2 \sigma ^2} ~~~{\rm with}~~~ 
    \frac{1}{a^2} = \frac{1}{\sigma ^2} - \frac{\beta}{2} ~.
\end{equation}
Note that both $\sigma$ and $\beta$ are involved in the normalization of $|\Psi _{test} \rangle$.
These test states are normalizable only when $\sigma < \sqrt{2 /\beta}$ due to the interplay between the fall off of the Gaussian test states and the exponential growth of $1/f(p)$ in the integral.
We will use this $\Psi (p)$ to calculate $\Delta p$ and $\Delta p '_p$, and then use these in \eqref{KMM-GUP1} to check if this leads to a minimum in $\Delta x '$. First, recall that $\Delta p ^2 = \langle p^2 \rangle - \langle p \rangle ^2$. Using the modified product we have $\langle p^2 \rangle = \int p^2 \Psi (p) ^* \Psi (p) e^{\beta p^2 /2} dp$ and $\langle p \rangle = \int p \Psi (p) ^* \Psi (p) e^{\beta p^2 /2} dp$. Using $\Psi (p)$ from \eqref{psi-test1} we find $\langle p^2 \rangle =  \frac{a^2}{2}$ and  $\langle p \rangle =  0$. Putting this together yields
\begin{equation}
    \label{deltap-g}
    \Delta p ^2 = \langle p^2 \rangle - \langle p \rangle ^2 = \frac{a^2}{2} =
    \frac{1}{2} \left( \frac{1}{\sigma ^2} - \frac{\beta}{2} \right)^{-1} ~.
\end{equation}
In the limit of ordinary quantum mechanics, $\beta \to 0$, we recover the usual result $\Delta p ^2 = \frac{\sigma ^2}{2}$. 

We now turn to calculate $\Delta {p'} _p ^2 = \langle {p'}^2 \rangle - \langle p ' \rangle ^2$ as a function of $\Delta p \sim \sigma$.  We recall that the expectations  $\langle {p'}^2 \rangle$ and $\langle p ' \rangle$ are carried out with respect to the modified measure {\it i.e.} $\langle p ' \rangle = \int p' \Psi (p) ^* \Psi (p) \frac{dp}{e^{-\beta p^2/2}}$ and $\langle {p'}^2 \rangle = \int {p'}^2 \Psi (p) ^* \Psi (p) \frac{dp}{e^{-\beta p^2/2}}$. It is easy to see from symmetry that, using $\Psi (p)$ from \eqref{psi-test1}, gives $\langle p ' \rangle = 0$. Next we calculate 
\begin{equation}
    \label{p2gauss}
    \langle {p'}^2 \rangle = \int _{-\infty} ^\infty \frac{dp}{e^{-\beta p^2 /2}} \frac{1}{\sqrt{\pi} a} e^{-p^2/2 \sigma ^2} \left( e^{\beta p^2/2} p \right) ^2  e^{-p^2/2 \sigma ^2}  = \frac{b^3}{2 a} ~,
\end{equation}
where $a$ is defined in \eqref{psi-test1} as $a = \left( \frac{1}{\sigma ^2} - \frac{\beta}{2} \right)^{-1/2}$ and $b$ is given by 
\begin{equation}
    \label{b}
    \frac{1}{b^2} = \left( \frac{1}{\sigma ^2} - \frac{3 \beta}{2} \right) ~~~ \to ~~~
    b = \left( \frac{1}{\sigma ^2} - \frac{3 \beta}{2} \right) ^{-1/2}
\end{equation}
Putting all this together gives $\Delta {p '}$ as
\begin{equation}
    \label{deltapp}
    \Delta {p'}_p = \sqrt{\langle {p'}^2 \rangle} = \sqrt{\frac{b^3}{2a}} = \frac{1}{\sqrt{2}} \left( \frac{1}{\sigma ^2} - \frac{3 \beta}{2} \right) ^{-3/4} \left( \frac{1}{\sigma ^2} - \frac{\beta}{2} \right)^{1/4} ~.
\end{equation}
Substituting $\Delta {p}$ from \eqref{deltap-g} and $\Delta {p'}_p$ from \eqref{deltapp} (and recalling that $\langle p' \rangle =0$) in equation \eqref{KMM-GUP1} give the following lower bound on $\Delta x'$:
\begin{equation}
\label{KMM-GUP2}
    \Delta  x'  \geq \frac{\hbar}{\sigma ^2 \sqrt{2}} \left( \frac{1}{\sigma ^2} - \frac{3 \beta}{2} \right) ^{3/4} \left( \frac{1}{\sigma ^2} - \frac{\beta}{2} \right)^{-5/4} ~.
\end{equation}

\begin{figure}
\centering
\includegraphics[width=125mm]{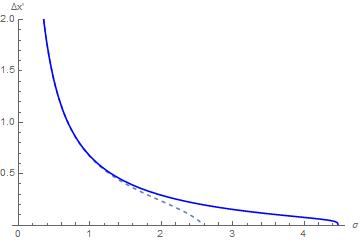}
\caption{$\Delta x '$ versus the spread in momentum $\Delta p \sim \sigma$ for the modified operators in \eqref{BAS-xp} with $\beta =0.1$. The dashed curved is $\Delta x '$ from the GUP using the equality in \eqref{KMM-GUP2} giving a lower bound. The solid curve is $\Delta x'$ calculated directly from the test wave function and the position operator in \eqref{dx-bas}}
\end{figure}

The $\Delta x'$ with the equality from \eqref{KMM-GUP2} is plotted as a function of width of the spread in the momentum, $\sigma$, via the dashed curve in Fig. 1 (in this and all plots we set $\hbar=1$). The dashed curve provides a lower limit to $\Delta x'$. We note that from \eqref{deltapp} and \eqref{KMM-GUP2}, one can see that $\Delta p'_p \to \infty$ and $\Delta x' \to 0$ as $\sigma \to \sqrt{2/3 \beta}$, which is less than $\sigma = \sqrt{2/\beta}$, the condition to have valid, normalizable states. Fig. 1 clearly shows that for the modified operators from \eqref{BAS-xp} there is no minimal length scale, in contrast to what happens for the modified position and momentum operators of reference \cite{KMM} in equation \eqref{KMM-xp}. This is despite the fact that both the modified operators in \eqref{KMM-xp} and those in \eqref{BAS-xp} give the same modified commutator in \eqref{GCOM}. 

We can also use the test wave function to directly calculate $\Delta x' = \sqrt{\langle {({\hat x'})}^2 \rangle - \langle {x'} \rangle ^2 } = \sqrt{\langle {({\hat x'})}^2 \rangle }$ where we have used that $\langle {\hat x'} \rangle =0$ for all the cases of modified position operator. 
For the modified position operator in \eqref{BAS-xp}, a straightforward calculation of  $\Delta x'$ with the test wave function \eqref{psi-test} gives
\begin{equation}
    \label{dx-bas}
    \Delta x' = \frac{\hbar}{\sigma ^2 \sqrt{2}} \left( \frac{1}{\sigma ^2} - \frac{\beta}{2} \right) ^{1/4} \left( \frac{1}{\sigma ^2} + \frac{\beta}{2} \right) ^{-3/4} ~.
\end{equation}
The plot of equation \eqref{dx-bas} is given by the solid curve in Fig. 1. As expected the shape of solid curve in Fig. 1 is similar to the dashed curve, with the dashed curve coming from saturation of the GUP in \eqref{KMM-GUP2}, providing a lower limit on $\Delta x'$. 

\subsection{Unmodified position and modified momentum}

Next we examine how the above issues play out for the modified operators from \eqref{BDV-1} where  ${\hat x}' = {\hat x} = i \hbar \partial_p$ and ${\hat p}' = p + \frac{\beta}{3}p^3$. 
Since, here ${\hat x}'$ is just the standard position operator, we use the standard inner product to maintain the symmetry of the operators. The constant $C$ in \eqref{psi-test} is the normalization factor with this standard inner product. Going through the standard calculation gives $C$ via $\int \Psi (p) ^* \Psi (p) dp =1 ~~\to~~ |C|^2 \int _{-\infty} ^{\infty} e^{-p^2 /\sigma^2} dp =1 ~~ \to ~~ C= \frac{1}{(\sqrt{\pi} \sigma) ^{1/2}}$. Thus in this case $|\Psi _{test} \rangle = \frac{1}{(\sqrt{\pi} \sigma) ^{1/2}} e^{-p^2 / 2 \sigma ^2}$. We now need to calculate $\Delta p$ and $\Delta p'_p$. Since ${\hat x}'$ is just the standard position operator we do not change the integration measure here and thus $\Delta p ^2 = \langle {\hat p}^2 \rangle - \langle {\hat p} \rangle ^2$ will just be the standard result for a Gaussian test wave function. Briefly, we find that $\langle {\hat p} \rangle =0$ and $\langle {\hat p}^2 \rangle = \frac{\sigma ^2}{2}$ for $|\Psi _{test} \rangle = \frac{1}{(\sqrt{\pi} \sigma) ^{1/2}} e^{-p^2 / 2 \sigma ^2}$. Thus $\Delta p ^2 = \langle {\hat p}^2 \rangle = \frac{\sigma ^2}{2}$, which is the standard result from standard quantum mechanics. Next $\Delta p' _p$ is given by 
\begin{equation}
    \label{ED-dpp}
    \Delta {p'}^2 _p = \left\langle \left({\hat p} +\frac{\beta}{3}{\hat p} ^3 \right) ^2  \right\rangle -
    \left\langle \left({\hat p} +\frac{\beta}{3}{\hat p} ^3 \right)  \right\rangle ^2 = \left\langle \left({\hat p}^2 +\frac{2\beta}{3}{\hat p} ^4 
    +\frac{\beta ^2}{9}{\hat p} ^6 \right)   \right\rangle ~.
\end{equation}
\begin{figure}
\centering
\includegraphics[width=125mm]{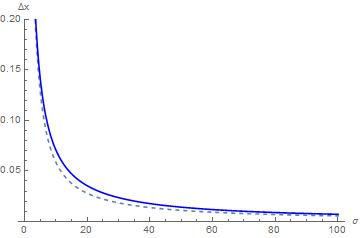}
\caption{$\Delta x$ versus $\Delta p' \sim \sigma$ from the modified operators from \eqref{BDV-1} for $\beta =0.1$. The dashed curve is $\Delta x$ obtained from the GUP, using the equality of \eqref{ED-dpp2}. The solid line is $\Delta x = \frac{1}{\sqrt{2}\sigma}$ from a direct calculation using the position operator in \eqref{BDV-1} and the test wave-function \eqref{psi-test}.}
\end{figure}
The second term in \eqref{ED-dpp} $\langle ( ...) \rangle ^2 =0$ by symmetry for $|\Psi _{test} \rangle = \frac{1}{(\sqrt{\pi} \sigma) ^{1/2}} e^{-p^2 / 2 \sigma ^2}$. The remaining term involves taking moments of ${\hat p} ^n$ with $n=2,4,6$ again using $|\Psi _{test} \rangle = \frac{1}{(\sqrt{\pi} \sigma) ^{1/2}} e^{-p^2 / 2 \sigma ^2}$. Carrying out the standard integrations gives
\begin{equation}
    \label{ED-dpp1}
    \Delta {p'}^2 _p = \left\langle \left({\hat p}^2 +\frac{2\beta}{3}{\hat p} ^4 
    +\frac{\beta ^2}{9}{\hat p} ^6 \right) \right\rangle = \frac{1}{2} \sigma ^2 + \frac{\beta}{2} \sigma ^4 + \frac{5\beta ^2}{24}  \sigma ^6~.
\end{equation}
Putting all this together in the uncertainty relation \eqref{KMM-GUP1} gives  
\begin{equation}
\label{ED-dpp2}
    \Delta  x  \geq \frac{\hbar}{2} \left( \frac{\left (1 + \beta (\Delta p)^2 - \beta \langle \hat p \rangle ^2 \right)}{\Delta p' _p} \right) = \frac{\hbar (1+\beta \sigma ^2 /2)}{\sigma \sqrt{2+2 \beta \sigma ^2 + 5 \beta ^2 \sigma ^4 /6}} ~.
\end{equation}
Since the position operator in this case is the standard one from quantum mechanics we have $\Delta x' = \Delta x$. In Fig. 2 we have plotted $\Delta x$ versus the width of the momentum, $\sigma$ for the equality in \eqref{ED-dpp2}. From Fig. 2 we see that for the modified operators from \eqref{BDV-1} there is no minimum in $\Delta x$, despite the fact that the modified operators from \eqref{BDV-1} lead to the same modified commutator as the modified operators from \eqref{KMM-xp}. As for the modified operators of the previous subsection, one can also calculate $\Delta x$ directly from the position operator in \eqref{BDV-1} and  the Gaussian test wave-function. Since here the position operator is just the usual one from standard quantum mechanics and the measure of the momentum integration is not altered, the spread in the position operator is just that of standard quantum mechanics for a Gaussian wave function, namely $\Delta x = \frac{1}{\sqrt{2} \sigma}$. This direct calculation of $\Delta x$ is plotted versus $\sigma$ by the solid curve. As previously we see that there is no minimum for $\Delta x$, and that the GUP calculated $\Delta x$ (the dashed curve) provides a lower bound on $\Delta x$.  

\subsection{Modified position and unmodified momentum}

In this subsection, as a check, we carry out the above procedure for the modified position and momentum operators from \cite{KMM} in equation \eqref{KMM-xp} for the test wave function $| \Psi _{test} \rangle = C e^{-p^2/2 \sigma}$. We find that for these modified operators one does obtain a non-zero minimum in agreement with the general analysis of \cite{KMM}. The normalization constant is determined by $\int _{-\infty} ^{\infty} \frac{dp}{1+\beta p^2} |C|^2 e^{p^2/\sigma ^2} = 1$, which gives $|C|^2 e^{1/\sigma ^2 \beta} \frac{\pi}{\sqrt{\beta}} {\rm Erfc}(1/\sqrt{\sigma ^2 \beta}) =1$, where ${\rm Erfc}$ is the complementary error function. Explicitly $C =  e^{-1/2 \sigma ^2 \beta} \left[ \frac{\pi}{\sqrt{\beta}} {\rm Erfc}(1/\sqrt{\sigma ^2 \beta}) \right] ^{-1/2}$. We want to calculate $\Delta {p'_p}^2 = \Delta p ^2 = \langle p^2 \rangle - \langle p \rangle ^2$. We have written $\Delta p' _p = \Delta p$ since, here, the momentum operator is just the same as in standard quantum mechanics. Now $\langle p \rangle = |C|^2 \int _{-\infty} ^{\infty} \frac{p e^{-p^2/\sigma ^2}}{1+ \beta p^2} dp$. By symmetry $\langle p \rangle = 0$ for  $| \Psi _{test} \rangle = C e^{-p^2/2 \sigma}$. For $\langle p ^2 \rangle$ we have
\begin{equation}
    \label{KMM-p1}
    \langle p ^2 \rangle = |C|^2 \int _{-\infty} ^{\infty} \frac{p^2 e^{-p^2/\sigma ^2} }{1+ \beta p^2}  dp = \frac{\sigma}{\sqrt{\beta \pi}} e^{-1/\sigma ^2 \beta}({\rm Erfc} (1/\sigma \sqrt{\beta} )^{-1} -\frac{1}{\beta}~,
\end{equation}
where we have used the explicit expression for $C$.
Equation \eqref{KMM-p1} gives $\Delta {p'_p} ^2 = \Delta p ^2 = \langle p ^2 \rangle = \frac{\sigma}{\sqrt{\beta \pi}} e^{-1/\sigma ^2 \beta}({\rm Erfc} (1/\sigma \sqrt{\beta} )^{-1} -\frac{1}{\beta}$. 
Since here the momentum operator is unaltered from the usual momentum operator from quantum mechanics we get $\Delta x'$ from \eqref{KMM-UP} as $\Delta x' = \frac{\hbar}{2} \left(\frac{1}{\Delta p} + \beta \Delta p \right)$. 
Using the expression for $\Delta p'_p = \Delta p$ that we obtained above from \eqref{KMM-p1} we obtain a not obviously enlightening expression for $\Delta x'$. We have plotted this $\Delta x'$ as a function of the spread in momentum , $\sigma$, in Fig. 3 with the dashed curve. Fig. 3 shows that in this case there is a minimum $\Delta x '$ in agreement with the general analysis from \cite{KMM}.  
\begin{figure}
\centering
\includegraphics[width=125mm]{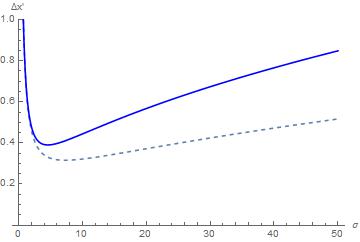}
\caption{$\Delta x '$ versus $\Delta p \sim \sigma$ for the modified operators in \eqref{KMM-xp} with $\beta =0.1$. The dashed curve is the GUP calculation of $\Delta x'$ and the solid curve is the direct calculation of $\Delta x'$.}
\end{figure}

One can also directly calculate $\Delta x'$ for the modified operators in \eqref{KMM-xp} for the Gaussian test wave function. 
A straightforward calculation for this case yields
\begin{equation}
    \label{direct-dx}
    \Delta x ' = \frac{1}{\sqrt{2}} \sqrt{\frac{1}{\sigma} + \frac{3 \beta \sigma}{2}} \left( e^{1/\sigma ^2 \beta} \sqrt{\frac{\pi}{\beta}} {\rm Efrc} (1/ \sigma \sqrt{\beta}) \right) ^{-1/2} ~.
\end{equation}
This direct calculation of $\Delta x'$ from \eqref{direct-dx} is plotted in Fig. 3 as the solid line. We see again that, as expected, there is a minimum $\Delta x'$ in this case. Also the GUP calculation of $\Delta x'$ in this case gives the lower limit on $\Delta x'$. 

\section{Summary and Conclusions}
 
We have shown that the existence of a minimal length in a GUP depends chiefly on how the position and momentum operators are modified rather than how the commutator is modified. 
Equations \eqref{KMM-xp}, \eqref{BAS-xp}, and \eqref{BDV-1} give various versions of modified position and momentum operators in momentum space which lead to exactly the same modified commutator in \eqref{GCOM}.
Only the modified operators in \eqref{KMM-xp} lead to a minimum length scale. 
The reason for this difference lies in equation \eqref{KMM-GUP1} which gives the relationship between position uncertainty, $\Delta x'$, and the momentum uncertainty $\Delta p' _p$. 
The numerator of \eqref{KMM-GUP1} always has the standard momentum uncertainty, $\beta \Delta p^2$, but the denominator varies greatly depending on how one modifies the momentum operator. 

The special feature of the modified position and momentum operators in \eqref{KMM-xp} that lead to a minimum non-zero length is that only the position operator is modified while the momentum operator is not. 
This means that $\Delta p'_p = \Delta p$ which makes the GUP of the modified operators in \eqref{KMM-xp} simple. 
In this case, $\Delta x'$ is bounded below by an expression of $\Delta p$ which has a strictly positive minimum. 
In the other cases, equations \eqref{BAS-xp} and \eqref{BDV-1}, the momentum operator is modified and it is not possible to obtain a general, analytical relationship between $\Delta p'_p$ and $\Delta p$. 
For the modified operators from \eqref{BAS-xp} and \eqref{BDV-1}, we selected a family of test states (Gaussians in momentum space with a width $\sigma$ given in equation \eqref{psi-test}) and then explicitly calculated the relationship between the uncertainty in position, $\Delta x'$, and the uncertainty in momentum characterized by the parameters $\sigma$ and $\beta$. 
If the test functions have no positive minimum for $\Delta x'$, then there is no minimum length scale for the position operator in general.  
It follows from these examples that the existence of such a minimum depends crucially on how one modifies the operators, more so than how the commutator is modified. Depending on the value of the parameter  $\beta$ one could potentially experimentally test for this quantum gravity motivated minimum length. A general approach to such experimental tests is given in the analysis of reference \cite{das1} using the Lamb shift, Landau levels, and  scanning tunnelling microscopy. In \cite{pikovski, chen} an experimental approach to testing for quantum gravity effects using quantum optomechanics is put forward. Quantum optomechanics is the use of light to prepare macroscopic mechanical system which is in a pure (or nearly pure) quantum state. A final, recent example of a concrete experimental proposal for testing quantum gravity can be found in references \cite{sujoy, sujoy1}. These two papers suggest a clever new experimental method of probing the parameter $\beta$ and minimum length scale using the (modified) rate of spreading of the wave-function of large molecules. This spreading of the wave-function method would provide a way to test the modification of the commutator as well as the modification of the operators.   

The results for the modified operators from \eqref{BAS-xp} and \eqref{BDV-1} are plotted in Figs. 1 and 2. There is no non-zero minimum in Figs. 1 and 2.
There is a significant difference between the results in Figs. 1 and 2: in Fig. 1,  $\Delta x'$ goes to zero in finite $\sigma$ while in Fig. 2, $\Delta x'$ goes to zero in the limit $\sigma \to \infty$.  
As a check, we applied the same test wave function to the modified operators of \eqref{KMM-xp}. 
The results are given in Fig. 3 and in this case do show a positive minimum in $\Delta x'$.
In section III, we used the GUP to determine the relationship between $\Delta x'$ and the uncertainty in the momentum using the test Gaussian wave function in \eqref{psi-test}. 
Due to the ``greater than or equal" nature of the GUP, this provides a lower limit on $\Delta x'$. 

All considerations in this work, of how quantum gravity might alter the standard Heisenberg uncertainty principle, involve only the spatial components of position and momentum. Recent work \cite{bosso} has explored the more complete space-time version of the position-momentum commutator. However, the general conclusions drawn are similar whether one uses the present version of the modified commutator or the space-time version of reference \cite{bosso}.

In conclusion, the specific modification of operators is crucial to obtaining a positive minimum length scale; the modified commutator is not generally sufficient for obtaining a minimum length scale.
From the analysis of \eqref{BDV-1}, the position operator needs to modified in order to obtain a minimum length scale.
For any unmodified position operator, the $\Delta x$ of the Gaussian test functions will go to zero as the width goes to infinity and that means there is no positive minimum length scale.
Whether one needs to modify the momentum operator to obtain a minimum length is not clear. 
The modified operators in \eqref{KMM-xp} lead to a minimum length without changing the momentum operator. 
The two examples of modified operators where the momentum operator is changed - equations \eqref{BAS-xp} and \eqref{BDV} - do not lead to a minimum length scale. However, careful analysis of \eqref{KMM-GUP1} may lead to examples where both the position and momentum operators are modified and give a minimum length scale. 
We can only say that we did not find a case where modifying the momentum operator led to a minimum length.   
This phenomenon, of the minimal length scale depending on the specific form of the modified operators, can be used to constrain the form of the operators one expects from quantum gravity. Can one modify the momentum operator and have a minimal length scale? If so, what forms can this modification take? What are the constraints on how the position operator is modified in order to obtain a minimum length scale? In future work, we plan to examine these constraints on the modification of the operators with the requirement that the modification leads to a minimal length scale.


\begin{thebibliography}{4}

\bibitem{maggiore}  M. Maggiore, Phys. Lett. B {\bf 304}, 65 (1993).

\bibitem{amati} D. Amati, M. Ciafaloni, and G. Veneziano, Phys. Lett. B {\bf 216}, 41 (1989).

\bibitem{amati2} D. Amati, M. Ciafaloni, and G. Veneziano, Phys. Lett. B {\bf 197}, 81 (1987); Int. J. Mod. Phys. A 03, 1615 (1988); Nucl. Phys. B {\bf 347}, 550 (1990).

\bibitem{gross} D.J. Gross and P.F. Mende, Phys. Lett. B {\bf 197}, 129 (1987); Nucl. Phys. B {\bf 303}, 407 (1988).

\bibitem{scardigli} F. Scardigli, Phys. Lett. B, {\bf 452}, 39 (1999).

\bibitem{garay} L.~J.~Garay, Int. J. Mod.\ Phys. A {\bf 10}, 145 (1995).

\bibitem{adler-1999} R. J. Adler, D. I. Santiago, Mod. Phys. Lett. A {\bf 14},  1371 (1999).

\bibitem{KMM} A. Kempf, G. Mangano and R. B. Mann, Phys. Rev. D {\bf 52}, 1108 (1995).

\bibitem{piero-2009} P. Nicolini, Int. J. Mod. Phys. A {\bf 24}, 1229 (2009).

\bibitem{hossenfelder} S. Hossenfelder, Living Rev.Rel. {\bf 16}, 2 (2013).

\bibitem{chang} L.~N.~Chang {\it et al.}, Adv. High Energy Phys.\ {\bf 2011}, 493514 (2011).

\bibitem{weinberg} S. Weinberg, {\it Lectures on Quantum Mechanics}, section 3.3 (Cambridge University Press, Cambridge 2013).

\bibitem{BAS} M. Bishop, E. Aiken, and D. Singleton, Phys. Rev. D {\bf 99}, 026012 (2019).

\bibitem{das2} A. Ali, S. Das, and E. Vagenas, Phys. Lett. B {\bf 678}, 497  (2009).

\bibitem{BDV} V. Balasubramanian, S. Das, and E. C. Vagenas, Annals Phys. {\bf 360}, 1 (2015). 

\bibitem{bosso1} P. Bosso, Phys.Rev. D {\bf 97}, 126010 (2018).

\bibitem{das1} S. Das and E. C. Vagenas,  Phys. Rev. Lett. {\bf 101}, 221301 (2008). 

\bibitem{pikovski} I.~Pikovski {\it et al.}, Nature Phys. {\bf 8}, 393 (2012).

\bibitem{chen} Y.~Chen, J. Phys. B {\bf 46}, 104001 (2013).

\bibitem{sujoy} C. Villalpando and S. K. Modak, Phys.Rev. D {\bf 100}, 024054 (2019).

\bibitem{sujoy1} C. Villalpando and S. K. Modak, Class.Quant.Grav. {\bf 36}, 215016 (2019).

\bibitem{bosso} V. Todorinov, P. Bosso and S. Das, Annals Phys. {\bf 405}, 92 (2019).

\end{thebibliography}
\end{document}